\documentclass{pasj00}

\begin{document}
\SetRunningHead{T. Shimizu, K. Masai, \& K. Koyama}{SNRs expanding out of the CSM into the ISM}
\Received{2011/07/19}
\Accepted{2011/10/05}

\title{Evolution of Supernova Remnants Expanding out of the Dense Circumstellar Matter into the Rarefied Interstellar Medium}

\author{Takafumi \textsc{Shimizu}\altaffilmark{1}, Kuniaki \textsc{Masai}\altaffilmark{1}, and Katsuji \textsc{Koyama}\altaffilmark{2} 
}

\altaffiltext{1}{Department of Physics, Tokyo Metropolitan University, 1-1 Minami-Ohsawa, Hachioji, Tokyo 192-0397}
\email{t-shimizu@phys.se.tmu.ac.jp}
\altaffiltext{2}{Department of Physics, Kyoto University, Kitashirakawa-Oiwake-cho, Sakyo-ku, Kyoto 606-8502}


\maketitle

\begin{abstract}

We carry out 3D-hydrodynamical calculations for the interaction of expanding supernova ejecta with the dense circumstellar matter (CSM) and the rarefied 
interstellar medium (ISM) outside. The CSM is composed of the stellar wind matter from the progenitor in its pre-supernova phase, and assumed to be axially symmetric: more matter around the equator than in the polar direction driven by rotation of the progenitor.  Because of high density of the CSM, the ionization state of the shock-heated ejecta quickly becomes equilibrium with the electron temperature.  When the blast wave breaks out of the CSM into the rarefied ISM, the shocked ejecta cools rapidly due to adiabatic expansion, and hence an over-ionized/recombining plasma would be left. 
The ejecta is reheated by the second reverse shock due to the interaction with the ISM.  We calculate the emission measure of the supernova remnant (SNR) along the line of sight, and find that the over-ionized plasma appears to be bar-like with wings in the edge-on (equatorial view), while shell-like in the face-on (polar view) geometry with respect to the rotation axis. 
The hot gas heated by the blast wave exists in the outermost region of the SNR with a nearly complete shell, but the X-rays therefrom are too faint to be observable. Thus, depending on the viewing angle, the SNR of the over-ionized plasma would exhibit center-filled morphology in X-rays, like W49B, a mixed-morphology SNR. The bar-like structure is swept out by the second reverse shock and disappears eventually, and then the SNR becomes shell-like in both the equatorial and polar views in the later phase of the evolution.

\end{abstract}

\section{Introduction}

Evolution of a supernova remnant (SNR) with no neutron star in its center is described essentially by the shock waves propagating outward (blast wave) into the ambient matter and inward (reverse-shock wave) into the supernova ejecta.  In young SNRs, the shocked matter is heated up to temperatures higher than $10^7$~K and forms an optically-thin hot plasma. This plasma emits X-rays with many emission lines of highly-ionized heavy elements.  Since the time-scale of ionization by electron impact is longer than that of  shock-heating of electrons,  X-ray spectra of young SNRs exhibit ionization states lower than those expected from the electron temperatures, which is called under-ionized or ionizing. The ionization state is represented by ionization temperature $T_{\mathrm{z}}$, the plasma temperature
in collisional ionization equilibrium having  the relevant ionization state. Thus, many of young SNRs show $T_{\mathrm{z}} < T_{\mathrm{e}}$ (under-ionized/ionizing) for the electron temperature $T_{\mathrm{e}}$. 

On the contrary to the above standard scenario, \citet{KO02} found an over-ionized plasma ($T_{\mathrm{z}} > T_{\mathrm{e}}$) in the {\it ASCA} observation from a mixed-morphology SNR, IC443. They proposed that thermal conduction from the hot interior of the remnant to the cold exterior can explain the over-ionized plasma. \citet{KO05} further investigated other five  mixed-morphology SNRs,  W49B, W28, W44, 3C 391 and Kes 27, and found that W49B shows $T_{\mathrm{z}} > T_{\mathrm{e}}$ as well.  

{\it Suzaku} detected clear spectral structures of radiative-recombination continua, direct evidence for recombining plasmas, in the X-ray spectra of IC443 \citep{YO09} and W49B \citep{OK09}.  Radiative recombination continua, and hence an over-ionized plasma is also found in another mixed-morphology SNR, G359.1--0.5 \citep{OK11}.  
Although the samples are still limited, the over-ionized plasmas are all found in the mixed-morphology SNRs associated in or near the star-forming complex with H\,II regions and molecular clouds. 
Hence these SNRs are likely due to core-collapse supernovae; the over-ionized plasma is possibly related to the massive 
progenitor and/or its environments. 

One possibility is that the ionization by photons of a few  tens keV at the initial phase of explosion of massive 
stars, e.g., possible X-ray flash or afterglow of $\gamma$-ray bursts, is responsible for the over-ionized state.  
Another possibility is rarefaction in the adiabatic evolution phase of the SNR;  when the shock wave breaks out of 
the dense circumstellar matter (CSM), like the progenitor's stellar wind, into the rarefied interstellar medium (ISM), 
the shock-heated electrons would rapidly cool due to adiabatic expansion, leaving highly ionized states \citep{itohmasai}.  

In the present paper, we investigate the latter possibility. \citet{itohmasai} calculated SNR evolution and 
radiation for a spherically symmetric structure.  We extend this work to non-symmetric structures as are found  
in mixed-morphology SNRs.  Unlike \citet{itohmasai}, we do not take ionization or radiation processes into account; 
we assume the relevant SNR evolution is fully adiabatic. 
We describe the model in the following section, calculation and results in Section 3, and discuss the results in 
Section 4. We give a summary in section 5 and notes about electron temperature in Appendix.  

\section{Model}

\subsection{Circumstellar matter}

In a core-collapse supernova, the massive progenitor (red or blue supergiant or Wolf-Rayet star) is expected to blow a strong stellar wind in its pre-supernova phase and forms dense CSM around the progenitor.  
If the stellar wind is spherically symmetric with a mass-loss rate $\dot{M}$ and wind velocity $v_{\mathrm{w}}$, the CSM density $\rho$ at a distance $r$ from the progenitor is given by

\begin{equation}
\rho=\frac{\dot{M}}{4\pi v_\mathrm{w}r^2}.
\end{equation}
 
\begin{figure}[htpb]
 \begin{center}
  \FigureFile(80mm,52mm){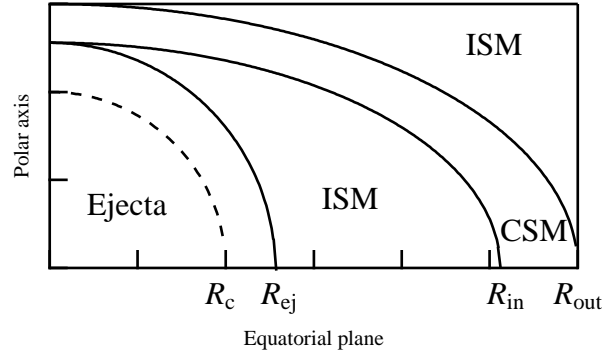}
\end{center}
\caption{Schematic picture of initial configurations for the interaction of the supernova ejecta with CSM.  
The broken and solid lines of the ejecta represent the radii of the core and envelope, respectively, 
and the outer two solid lines represent the inner and outer boundaries of the CSM to the ISM.}
\label{figure1}
\end{figure}

For a while before supernova explosion, the stellar wind ceases. Then the stellar wind forms a thick shell. 
The inner and outer radii of the shell are given by $R_{\mathrm{in}}=v_\mathrm{w}t_\mathrm{e}$ and 
$R_{\mathrm{out}}=v_\mathrm{w}(t_\mathrm{w}+t_\mathrm{e})$, respectively (see Figure \ref{figure1}),
where time $t_\mathrm{w}$ and $t_\mathrm{e}$ are the duration of the wind activity and the elapsed time after the cease of the wind, respectively.
  
In realistic stellar winds,  more matter around the equatorial plane than the polar direction may be accumulated, because of the rotation of the progenitor; disk-like winds of Be stars for an extreme example.  
We thus modify Eq. (1) as,  
\[\rho=\frac{\dot{M}}{4\pi v_\mathrm{w}}\frac{a}{x^2+y^2+a^2z^2},\]
\begin{equation}
R_{\mathrm{in}}\leq (x^2+y^2+a^2z^2)^{1/2} \leq R_{\mathrm{out}}
\end{equation}
with an anisotropy parameter $a$, which is the ratio of the major to minor axis of an equi-density surface of the wind matter.  
Hereafter, $R_{\mathrm{in}}$ and $R_{\mathrm{out}}$ are referred to the radii on the equatorial plane.  
This distribution gives $a^2$ times higher density on the equatorial plane than in the polar direction at the same distance.  For the progenitor of SN~1987A, \citet{BL93} suggested that the ratio of equatorial to polar mass-loss rate was at least 20 during its red supergiant stage. This value corresponds to $a \simeq 4.5$.  In our calculations we adopt the values from 1 to 3.

We calculate the evolution for the following five models: 
\begin{itemize}
\item[A1] Isotropic CSM same as \citet{itohmasai} for a reference.
\item[A2] Anisotropic CSM concentrated around the equatorial plane.
\item[A3] Same as A2, but more mass around the equatorial plane.
\item[B1] Same as A2, but with a lower density ISM.
\item[B2] Same as B1,  but with a higher wind velocity.
\end{itemize}
The numerical values for each model are shown in Table \ref{table1}, where values of $R_{\mathrm{in}}$, $R_{\mathrm{out}}$, $\rho_{\mathrm{in}}$ and $\rho_{\mathrm{out}}$ are those on the equatorial plane.  The mass-loss rate is assumed to be 
$5 \times 10^{-5}\MO \mbox{ yr}^{-1}$, the same as \citet{itohmasai}, for all the models. 

\begin{table*}
\caption{Parameters of circumstellar matter and interstellar medium}
\begin{center}
\begin{tabular}{cccccccc}
\hline
&\multicolumn{6}{c}{CSM}&ISM\\
\hline
Model&$a$&$R_{\mathrm{in}}$ (10$^{16}$cm)&$R_{\mathrm{out}}$ (10$^{16}$cm)&$\rho_{\mathrm{in}}$ (amu$\mbox{ cm}^{-3}$)&$\rho_{\mathrm{out}}$ (amu$\mbox{ cm}^{-3}$)&$v_{\mathrm{w}}$ (cm$\mbox{ s}^{-1}$)&$\rho_{\mathrm{ISM}}$ (amu$\mbox{ cm}^{-3}$)\\
\hline
A1&1   &2.0&92.6&3.4$\times10^{5}$   &1.6$\times10^{2}$&10$^6$         &1.6$\times10^{-1}$\\
A2&2   &2.0&92.6&6.9$\times10^{5}$   &3.2$\times10^{2}$&10$^6$         &1.6$\times10^{-1}$\\
A3&3   &2.0&92.6&1.0$\times10^{6}$   &4.8$\times10^{2}$&10$^6$         &1.6$\times10^{-1}$\\
B1&2   &2.0&92.6&6.9$\times10^{5}$   &3.2$\times10^{2}$&10$^6$         &1.6$\times10^{-2}$\\
B2&2   &20&926&6.9$\times10^{2}$   &3.2$\times10^{-1}$&10$^7$&1.6$\times10^{-2}$\\
\hline
\multicolumn{8}{l}
{\small Mass loss rate $\dot{M}$ is $5\times10^{-5}\MO \mbox{ yr}^{-1}$ in the all models, and $\rho_{\mathrm{in}}=\rho(r=R_{\mathrm{in}})$, $\rho_{\mathrm{out}}=\rho(r=R_{\mathrm{out}})$.}\\
\end{tabular}
\end{center}
\label{table1}
\end{table*}

\subsection{Supernova ejecta}
We focus on the effect of the CSM, particularly for anisotropic CSM, on the early phase evolution of SNRs. For the supernova  
ejecta, we simply assume spherical distribution with a core of radius $R_{\mathrm{c}}$ and extended envelope of radius $R_{\mathrm{ej}}$, as \citep{TM99}
\[
\rho= \frac{3M_{\mathrm{ej}}}{4\pi v_{\mathrm{ej}}^3t^3}\frac{1-n/3}{1-(n/3)w_{\mathrm{c}}^{3-n}}~\times
\]
\begin{equation}
\left\{\begin{array}{cc} w_{\mathrm{c}}^{-n} & \mbox{for } r < R_{\mathrm{c}}\\
	   			      (r/({v_{\mathrm{ej}}t}))^{-n} & \mbox{for } R_{\mathrm{c}} \leq r \leq R_{\mathrm{ej}} \end{array}\right.,
\end{equation}
where
\begin{equation}
v_{\mathrm{ej}}=\left(\frac{2E_{\mathrm{ej}}}{M_{\mathrm{ej}}}\right)^{1/2}\left(\frac{5-n}{3-n}\right)^{1/2}
\left(\frac{w_c^{-(3-n)}-n/3}{w_c^{-(5-n)}-n/5}\right)^{1/2}\frac{1}{w_c}
\end{equation}
is the expansion velocity at $R_{\mathrm{ej}}$.  
$M_{\mathrm{ej}}$ and $E_{\mathrm{ej}}$ are the ejecta mass and the explosion energy (kinetic energy), respectively. 
These are assumed  to be $E_{\mathrm{ej}}=2\times10^{51}$ erg and  $M_{\mathrm{ej}}=10\MO$.
The other parameters $n$ and $w_{\mathrm{c}}=R_{\mathrm{c}}/R_{\mathrm{ej}}$ are taken to be 6 and 0.49, respectively.
All these values  are the same as \citet{itohmasai}, and hence  $v_{\mathrm{ej}} \simeq 8.5\times10^8$ cm$\mbox{ s}^{-1}$. 

\section{Calculation and Results}

We numerically solve the Euler equations utilizing the athena3d code (\cite{stone}): 
\begin{eqnarray}
&&\frac{\partial \rho}{\partial t}+\nabla\cdot(\rho\boldsymbol{v})=0,\nonumber\\
&&\frac{\partial \boldsymbol{v}}{\partial t}+(\boldsymbol{v}\cdot\nabla)\boldsymbol{v}=-\frac{1}{\rho}\nabla p,\\
&&\frac{\partial}{\partial t}\{\rho(\frac{1}{2}v^2+u)\}+\nabla\cdot\left\{\rho\boldsymbol{v}\left(\frac{1}{2}v^2+u+\frac{p}{\rho}\right)\right\}=0,\nonumber
\end{eqnarray}
where $\rho$, $p$, $\boldsymbol{v}$ and $u$ are the mass density, the pressure,  the fluid velocity and the internal energy per unit mass, 
respectively.  The equation of state is written as $p=(\gamma-1)\rho u$ with $\gamma=5/3$, and the mean molecular weight is taken to be 0.5.  

In the hydrodynamical evolution,  we obtained mean temperature $T$ of ions (protons) and electrons weighted by their number density. If no plasma mode works directly on electrons at the shock front, the ion temperature $T_{\mathrm i}$ rises faster and then the energy of ions is transferred to electrons. 
We calculate  $T_{\mathrm i}$ and $T_{\mathrm e}$ from $T$, assuming the transport by Coulomb collisions (\cite{masai94}; see Appendix). 

\subsection{Model A1}

Although  we include  neither ionization nor radiation and simply assume the adiabatic evolution, 
the evolution of model A1 is in fairly good agreement with \citet{itohmasai}.  
The blast shock and reverse shock heat the CSM to $T \gtrsim 10^8$~K, and  the ejecta to $T \gtrsim 10^7$~K, respectively.  
The blast wave expands as $r_{\mathrm{b}}\propto t^{0.88}$, while 
$r_{\mathrm{b}}\propto t^{0.87}$ in \citet{itohmasai}, where $r_{\mathrm{b}}$ is the blast-wave radius. 

The blast wave breaks out at 41~yr, 2 years earlier than in \citet{itohmasai} who take radiation loss into calculation. 
After the break-out, the blast wave rapidly expands adiabatically, and hence the temperature and density of the shocked matter decrease approximately as $T \propto t^{-2}$ and $\rho \propto t^{-3}$ for $\gamma = 5/3$. 

When the pressure of the shocked CSM becomes below that of the shocked ISM, the second reverse shock occurs to propagate 
inward and reheats the CSM and ejecta. The second reverse shock reaches the ejecta at 830~yr, 120 years earlier 
than in \citet{itohmasai}.

\subsection{Models A2 and A3}
\begin{figure}[htpb]
 \begin{center}
  \FigureFile(80mm,78mm){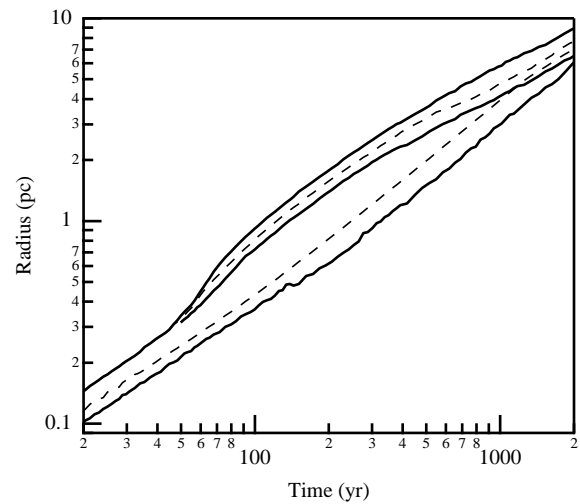}
 \end{center}
\caption{Shocks in the equatorial direction of model A2 before and after the blast-wave break-out, as functions of the 
elapsed time after explosion.  The upper, middle and lower solid lines represent the radii of the blast wave, second 
reverse shock and reverse shock, respectively.  The upper and lower broken lines represent the outer radii of the CSM 
and ejecta, respectively.}
\label{figure2}
\end{figure}

The evolution is basically the same as model A1, but is dependent on direction: more matter on the equatorial plane than in the polar direction. With increasing the value of $a$, the break-out occurs earlier in the polar and later in the equatorial direction than in model A1. The former is due to a shorter distance to the CSM, and the latter is higher density of the CSM than those in model A1. 
In addition, the second reverse shock reaches the ejecta earlier in the polar direction and later in the equatorial direction than in 
model A1.  In model A2, the break-out occurs at $17$~yr in the polar direction, and at $46$~yr in the equatorial direction as shown in 
Figure \ref{figure2}.  
The mean temperature $T$ of the shocked ejecta turns to rise at 340~yr and 1100~yr,
in the polar and equatorial directions, respectively.

\subsection{Model B1}
The break-out occurs at the same time as model A2 since the CSM distribution is the same as model A2. 
The difference of model B1 from model A2 is the lower density of the ISM by an order of magnitude.  The velocity of 
the blast wave after the break-out is slightly higher than that in model A2 by a factor of $\sim1.4$ in the equatorial 
direction.  As a result, the pressure of the shocked ISM is about one-fifth of that in model A2, and the second reverse shock 
occurs later. The temperature $T$ of the shocked ejecta drops at 23~yr and 68~yr by the break-out, but turns to rise at 810~yr and  1800~yr in the polar and equatorial directions, respectively.

\subsection{Model B2}
The difference from model B1 is the lower density of the CSM by three orders of magnitude due to higher wind velocity 
by an order of magnitude. Even at such low densities, the break-out occurs at 170~yr and  450~yr
in the polar and equatorial directions, respectively.  Similarly to the other models, the temperature $T$ of the  
shocked ejecta drops at 280~yr and 770~yr, but turns to rise by the second reverse shock at 920~yr and  2400~yr in the polar and  equatorial directions, respectively.

\section{Discussion}

\begin{table*}
\caption{Characteristic Epochs}
\begin{center}
\begin{tabular}{ccccccc}
\hline
&\multicolumn{3}{c}{Break-out}&\multicolumn{3}{c}{$T_{\mathrm{z}}-T_{\mathrm{e}}$ Decoupling}\\
\hline
Model&$t_{\rm b}$ (yr)&$T$ ($10^7$K)&$R_{\mathrm{b}}$ (pc)&$t_{\rm d}$ (yr)&$T$ ($10^7$K)&$R_{\mathrm{b}}$ (pc)\\
\hline
A1&41&1.3&0.30&180&0.20&1.7\\
\hline
A2-e&46&1.2&0.30&210&0.18&1.8\\
A2-p&17&1.6&0.15&80&0.16&0.99\\
\hline
A3-e&47&1.4&0.30&220&0.21&1.8\\
A3-p&11&0.8&0.10&60&0.11&0.81\\
\hline
B1-e&46&1.2&0.30&190&0.23&2.0\\
B1-p&17&1.7&0.15&80&0.10&1.1\\
\hline
B2-e&450&1.2&3.0&70&1.7&0.64\\
B2-p&170&1.1&1.5&36&1.2&0.38\\
\hline
\multicolumn{7}{p{11cm}}
{The characters ``e" and ``p" attached to Model A2--B2 mean the equatorial and polar directions, respectively.}\\
\end{tabular}
\end{center}
\label{table2}
\end{table*}

In Table \ref{table2} we summarize the temperatures and blast-wave radii at characteristic epochs, $t_{\rm b}$ and $t_{\rm d}$ (see below), for each model. 

\subsection{Dynamical evolution}

\begin{figure}[htpb]
\begin{center}
\FigureFile(80mm,65mm){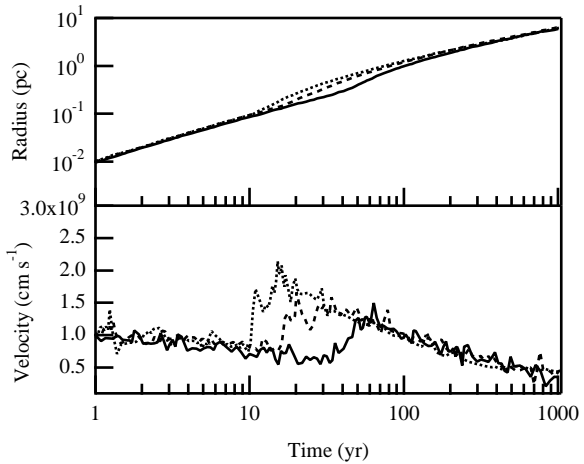}
\end{center}
\caption{Radius and velocity of the blast wave in the polar direction for models A1 (solid lines), A2 (dashed lines) 
and A3 (dotted lines), as functions of elapsed time after explosion.}
\label{figure3}
\end{figure}

Before the break-out, the evolution of SNR can be described by a self-similar solution, and the blast wave expands 
as $r_{\mathrm{b}}\propto t^{\alpha}$. Almost independently of  anisotropy of the CSM (models A1--3), the values of 
$\alpha$ are close to each other both in the polar and equatorial directions, as seen in the {\it upper} panel 
of Figure \ref{figure3}.  After the break-out, $\alpha$ values are slightly larger in larger $a$ models, but approach the value of model A1 ($a = 1$) with time.  This implies that the blast wave approaches spherical symmetry with time, and the ratio of the blast-wave radius of the polar to equatorial direction becomes less than $\sim1.2$ in 1000 years in any model of ours here. 

If the density difference across the interface between the CSM and ISM is large enough 
(models A1--3 and B1), the blast wave gets faster just after the break-out by a factor of $\sim 2$,
almost independent of the density difference.  
On the other hand, if the density difference is as small as model B2, the velocity increases by a factor of 
$\sim 1.4$.  The expansion velocity of a spherical fluid initially at rest asymptotically approaches its maximum value of 
$2c_{\mathrm{s}}/(\gamma-1)$, where $c_{\mathrm{s}}$ is a sound speed of the fluid \citep{Zel'dovich66}. 
Similarly, denoting the shock velocity by $V_{\mathrm S}$, we may have the maximum expansion velocity after the break-out, as
\begin{equation}
\frac{2}{\gamma-1}[c_{\mathrm s}^2+V_{\mathrm S}^2(t_{\mathrm b})]^{1/2}\simeq 2.0\,V_{\mathrm S}(t_{\mathrm b}),
\end{equation}
where $V_{\mathrm S}(t_{\mathrm b})$ is the shock velocity of the blast wave immediately before the break-out at $t = t_{\mathrm b}$. Here we take the Rankine-Hugoniot relation of strong shocks into account. This estimate is consistent with the result of our hydrodynamical 
calculation. The high shock velocity and rarefaction caused by the break-out may be in favor of particle acceleration, because the 
former boosts the maximum energy of particles, and the latter works for particles to become non-thermal.   

\subsection{Temperature and Ionization state}

\begin{figure*}[htpb]
\begin{center}
\FigureFile(80mm,70mm){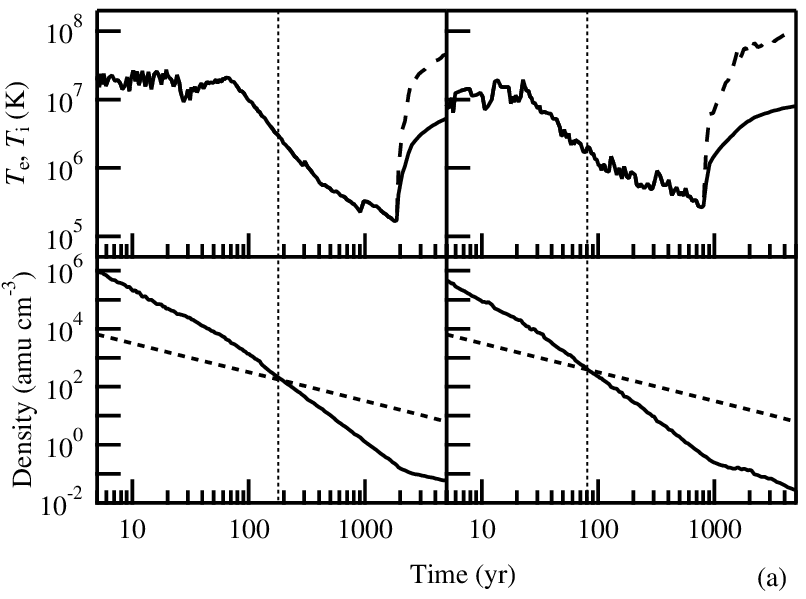} \FigureFile(80mm,70mm){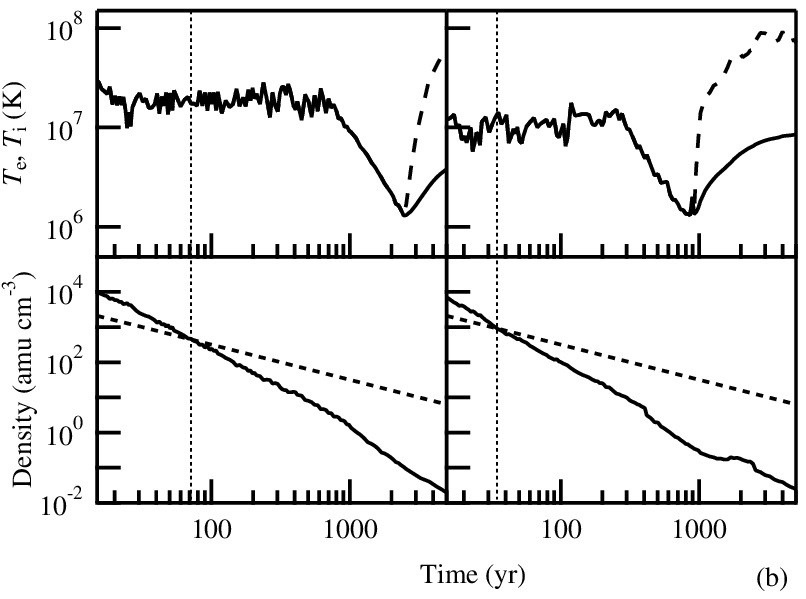}
\end{center}
\caption{Model (a) B1 and (b) B2: Averaged temperature and density of the shocked ejecta in the equator ({\it left}) and polar ({\it right}) 
directions, as functions of the elapsed time after explosion. {\it Upper}: The solid and broken lines 
represent the electron and ion temperatures, respectively.  {\it Lower}: The solid and dashed lines represent the density and $\rho_{\rm crit} = 10^{12}\,t^{-1}$~amu~cm$^{-3}$, respectively, and the thin-dotted vertical lines represent the decoupling epoch $t_{\rm d}$ (see text).}
\label{figure4}
\end{figure*}

As demonstrated by \citet{itohmasai}, rarefaction by the blast-wave break-out rapidly cools the shock-heated matter to be an over-ionized/recombining plasma of $T_{\mathrm e} < T_{\mathrm z}$. This recombining plasma state lasts until the second reverse shock reheats the matter to $\gtrsim T_{\mathrm z}$.  Although ionization is not calculated simultaneously with hydrodynamics, we can analyze the ionization state, i.e., ionizing, equilibrium or recombining, of the shocked matter from the electron temperature and density. The ionization state thus evaluated is consistent with \citet{itohmasai}, considering the variation of epochs and duration that depend on the CSM/ISM models, as discussed in the previous sections.

The {\it upper} panels of Figures \ref{figure4}a (model B1) and \ref{figure4}b (model B2) show the evolution of the ion and electron temperatures averaged over the shocked matter of the ejecta. Before the break-out, $T_{\mathrm e}$ becomes nearly equal to $T_{\mathrm i}$ due to high densities of the CSM.  Also after the break-out and until arrival of the second reverse shock, $T_{\mathrm e}$ is nearly equal to $T_{\mathrm i}$, because cooling is due to adiabatic expansion. By the second reverse shock, $T_{\mathrm i}$ rises faster than $T_{\mathrm e}$, and then $T_{\mathrm e}$ rises through 
the energy transport from ions, as shown in the {\it upper} panels of Figure \ref{figure4}. 

In the {\it lower} panels, we show the average density of the shocked ejecta by the solid line. Ionization by electron-impact becomes equilibrium with time-scale $\tau$ given by $\rho \tau \sim 10^{12}\mbox{ amu cm}^{-3}\mbox{ s}$, 
almost independent of the temperature or ion species \citep{masai94}. We also plot this relation as $\rho_{\rm crit} = 10^{12}\,t^{-1} \mbox{ amu cm}^{-3}$ with the dashed line in the {\it lower} panels. Well before the break-out, since the density is high enough ($\rho > \rho_{\rm crit}$) due to the presence of CSM, ionization quickly reaches its equilibrium at $T_{\mathrm e}$, i.e. $T_{\mathrm z} \sim T_{\mathrm e}$. 

As the SNR expands, the average density of the shocked matter decreases approximately as $\rho \propto t^{-2}$.
When rarefaction occurs by the break-out, 
$\rho$ decreases faster as $\propto t^{-3}$ (see Section 3.1). 
Therefore, $\rho$ becomes below $\rho_{\rm crit} \propto t^{-1}$, 
as seen in the {\it lower} panels of Figure \ref{figure4},  
and $T_{\mathrm z}$ is decoupled from $T_{\mathrm e}$. 
In other words, recombination no longer follows the rapid decrease of the electron temperature, and the ionization state freezes roughly at $T_{\mathrm z} \sim T_{\mathrm e} (t_{\rm d})$, where $t_{\rm d}$ is the epoch at which $\rho = \rho_{\rm crit}$. 

The decoupling epoch $t_{\rm d}$ is also shown in the {\it lower} panels of Figure \ref{figure4} with the thin-dotted vertical line. In model B1, $t_{\rm d} \sim 190$~yr in the equatorial direction.  Note that $t_{\rm d} > t_{\rm b}$ and $T_{\mathrm e}(t_{\rm d})<T_{\mathrm e}(t_{\mathrm b})$ in model B1, where $T_{\mathrm e}(t_{\mathrm b})$ is the electron temperature immediately before the break-out.  Models A1--3 show the similar behavior to model B1 described here. 

In model B2, unlike other models or \citet{itohmasai}, the CSM is located away from the progenitor, and hence the density is low.  Therefore, the break-out occurs later but $\rho$ becomes below $\rho_{\rm crit}$ earlier than in model B1.  As seen in Figure \ref{figure4}b, $t_{\rm d} < t_{\rm b}$ in model B2, and $t_{\rm d} \sim 70$~yr in the equatorial direction. 
In fact, as is seen in the {\it upper} panel, $T_{\mathrm e}$ before the break-out is nearly constant at $\sim T_{\mathrm e}(t_{\mathrm b})$ in model B2.  As a result, the ionization temperature of the recombining plasma is higher in model B2 than that in model B1. 

When the second reverse shock reaches the ejecta, the temperature turns to rise, as seen in the {\it upper} panels of Figure \ref{figure4}. If the electron temperature exceeds or becomes comparable to $T_{\mathrm z} \sim T_{\mathrm e}(t_{\rm d})$, the ionization state turns to be ionizing or nearly equilibrium.  This is the case for models A1--3 with a higher ISM density.  In model B1, though the ISM density is low, $T_{\mathrm e}$ exceeds $T_{\mathrm z}$ in 3000 years owing to the dense CSM and lower $T_{\mathrm z}$. On the other hand, in model B2, the density is too low for the second reverse shock to raise $T_{\mathrm e} > T_{\mathrm z}$ in such short time, and the over-ionization state remains much longer. If $T_{\mathrm i}$ becomes higher than $2T_{\mathrm z}$ at the shock front, $T_{\mathrm e} \gtrsim T_{\mathrm z}$ may be attained by the energy transfer from ions.  Since $T_{\mathrm i} \propto V_{\mathrm S}^2$ (reverse shock) decreases with the age, however, the recombining state would last yet for thousands years. 

After decoupled from the electron temperature, the ionization temperature decreases by recombination with time-scale $\tau$.  We note here that the recombination time-scale $\sim 10^4$ yr \citep{masai94} is longer than the age of over-ionized  SNRs: $\sim 4000$~yr for IC443 \citep{Troja08} and $1000 - 4000$~yr for W49B (\cite{Pye84}, \cite{Smith85}, Hwang et al. 2000).

\subsection{X-ray emission measure distribution}

\begin{figure}[htpb]
\begin{center}
\FigureFile(84mm,46.2mm){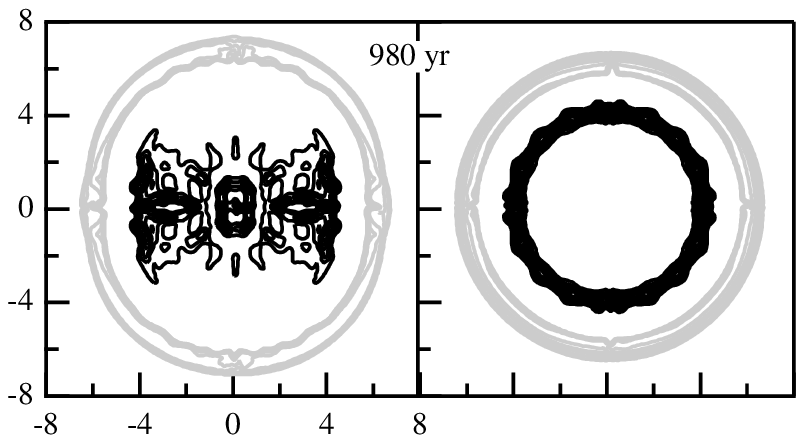} \\
\FigureFile(84mm,46.2mm){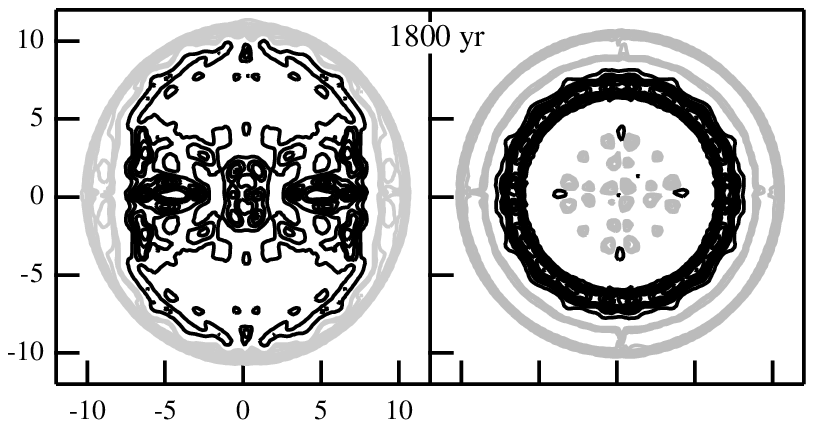} \\
\FigureFile(84mm,46.2mm){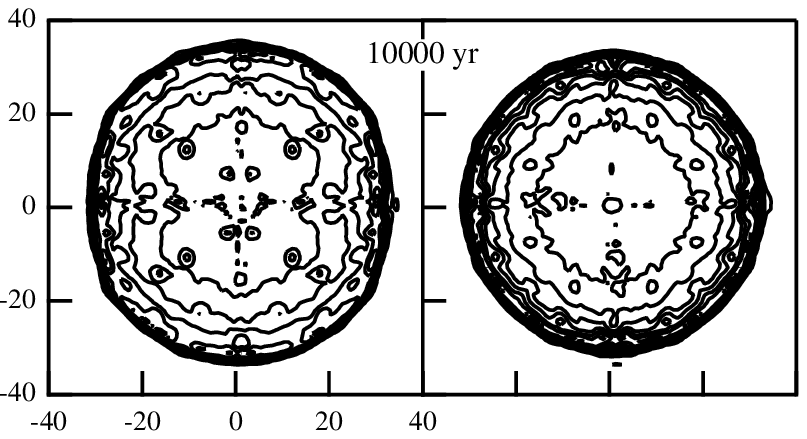}
\end{center}
\caption{Contours of the X-ray emission measure in the equator ({\it left}) and polar ({\it right}) directions of model B2 are drawn by black lines linearly from zero, every $1 \times 10^{19} \mbox{ amu}^2 \mbox{ cm}^{-5}$ to the maximum $\sim 1.3 \times 10^{20} \mbox{ amu}^2 \mbox{ cm}^{-5}$ at 980~yr, every $5 \times 10^{17} \mbox{ amu}^2 \mbox{ cm}^{-5}$ to the maximum $\sim 3.8 \times 10^{18} \mbox{ amu}^2 \mbox{ cm}^{-5}$ at 1800~yr, and every $2 \times 10^{16} \mbox{ amu}^2 \mbox{ cm}^{-5}$ to the maximum $\sim 1.9 \times 10^{17} \mbox{ amu}^2 \mbox{ cm}^{-5}$ at 10000~yr, after explosion.  The grey contours represent the low-level emission measure drawn linearly from zero, every $1 \times 10^{16} \mbox{ amu}^2 \mbox{ cm}^{-5}$ up to $1.0 \times 10^{17} \mbox{ amu}^2 \mbox{ cm}^{-5}$ at 980~yr, every $1 \times 10^{16} \mbox{ amu}^2 \mbox{ cm}^{-5}$ up to $1.0 \times 10^{17} \mbox{ amu}^2 \mbox{ cm}^{-5}$ at 1800~yr.  The horizontal and vertical axes show the scale in units of pc.}
\label{figure5}
\end{figure}

All the SNRs with over-ionized/recombining plasma so far observed are mixed-morphology SNRs, which exhibit center-filled X-ray emission and shell-like radio emission.  Hence, we investigate distribution of the X-ray emission measure for SNRs with anisotropic stellar wind. We integrate the density square along the line of sight, as $\int \rho^2 dl$, and show the map (black lines) for model B2 in Figure \ref{figure5}.  Here the shocked matter of $T > 10^6$~K, which is responsible for X-ray emission, is taken into calculations.

At ages 980~yr and 1800~yr, we can see a bar-like structure (horizontal) with diffuse wings in the equatorial view, while a thin shell in the polar view.  The emission measure is dominated by the shock-heated ejecta, which is a recombining plasma as seen from Figure \ref{figure4}b. 
The temperature dependence of the line emissivity is weaker in recombining than in equilibrium, and the emission-measure map reflects roughly the X-ray surface brightness. 

The bar structure is clearer at younger ages after the break-out, and diffuses out gradually with time.  At 1800~yr, the second reverse-shock front appears to surround the bar/wing structure, and thereby the bar-end structure is being distorted. 
The second reverse-shock propagates inward and eventually sweeps the bar/wing structure out.  Hence, the late-phase SNR approaches shell-like both in the equatorial and polar views, as seen at 10000~yr in Figure \ref{figure5}, where a little elongated shape with a narrow middle part is still seen in the equatorial view. 

W49B shows a bar-like structure as well as a recombining plasma state \citep{OK09}. The bar and its east-end observed with {\it Chandra} \citep{Keohane07} look similar to our calculation shown in Figure \ref{figure5}. Thus we suggest that W49B is the case of nearly equatorial view for the supernova remnant of a massive progenitor which exploded in its past stellar wind. It should be noted that, in our calculation, the structure of the bar-end is formed by the reverse shock, not by the collision with a  molecular cloud near the bar-end.   

The grey lines in Figure \ref{figure5} represent the low-level emission measure, much smaller than that represented by the black lines. One can see the blast-wave front in the grey lines.  The shocked matter in the grey lines is hardly observed in X-rays, but the shock possibly accelerates electrons to be of order of GeV, which can be responsible for GHz synchrotron radio; the condition after the break-out is favorable for particle acceleration, as mentioned in Section 4.1. 

\section{Conclusion}
We investigate the evolution of SNRs 
that explode in the progenitors' stellar wind matter, 
considering possible environments of mixed-morphology SNRs with over-ionized plasmas.  
We summarize the results;

\begin{itemize}
\item When the blast wave breaks out of the wind matter into the ambient interstellar medium, the shocked matter cools rapidly due 
to adiabatic expansion.  Just after the break-out, the expanding velocity becomes faster by a factor up to two, and then gradually decreases to that of the extrapolated from the velocity trend before the break-out.

\item Before the break-out, the shocked matter reaches ionization equilibrium and equipartition $T_{\mathrm{e}} \sim T_{\mathrm{i}}$, but deviates from equilibrium by rarefaction after the break-out.  Consequently, the shock-heated ejecta turns to be a recombining plasma, since cooling due to adiabatic expansion is much faster than recombination.

\item The recombining state of the shocked ejecta lasts until the second reverse shock, which occurs by the interaction with the interstellar medium, propagates inward and reheats the ejecta.  If the density of the ejecta is too low to establish ionization equilibrium, however, the recombining state lasts longer.

\item After the break-out in the adiabatic phase, since the emission measure of the shocked ejecta is much larger than that of the shocked ambient matter, the SNR in X-ray wavelengths appears much brighter in the reverse-shocked inner region than the blast-shocked outer shell. 

\item When the stellar wind matter is not isotropic but denser in the equatorial direction 
due to the progenitor's rotation, the SNR in the recombining state looks bar-like with wings in the equatorial view and thin shell-like in the polar view. So that, the SNR would show center-filled various shapes in X-rays, depending on the viewing angle.  On the other hand, the blast-shocked matter, which is very faint in X-rays but could be observed in radio, forms a fairly complete shell outside. 

\item As the SNR age increases, however, the second reverse shock sweeps the bar/wing structure out and merges into the whole ejecta eventually.  Hence, the bar/wing structure disappears and the late-phase SNR would look shell-like almost independently of the viewing angle. 

\end{itemize}

\bigskip 
The authors are grateful to Hiroya~Yamaguchi and Midori~Ozawa for meaningful discussion about {\it Suzaku} observations of 
recombination radiation from SNRs.  KM and KK were respectively supported by the Grant-in-Aid for
Scientific Research 22540253 and 2054019, from Japan Society for the Promotion of Science (JSPS).

\bigskip

\appendix
\section{Electron temperatures of SNRs}
\begin{figure}[htpb]
\begin{center}
\FigureFile(82.9mm,63.1mm){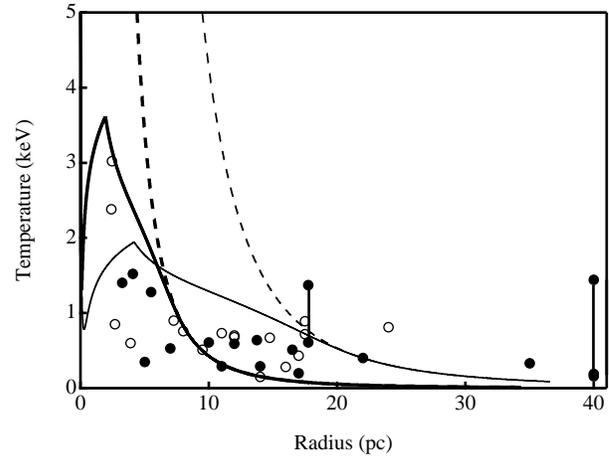}
\end{center}
\caption{Electron temperature vs. radius of SNRs which have no active central sources.  Open and filled circles represent shell-like and mixed-morphology SNRs, respectively; line-connected two circles mean the data by two-temperature analysis.  The thin and thick solid/broken lines represent the calculations for the density $\rho = 0.1 \mbox{ amu cm}^{-3}$ and $\rho = 1 \mbox{ amu cm}^{-3}$, respectively, of the ambient matter. 
\newline
References.---Kepler; \cite{KT99},
G15.9+0.2; \cite{Reynolds06},
G27.4+0.0; \cite{GV97},
Cygnus Loop; \cite{Miyata07},
G109.1+1.0; \cite{Sasaki04},
Cassiopeia A; \cite{KO05},
Tycho; \cite{Hwang97},
G156.2+5.7; \cite{Katsuda09},
Puppis A; \cite{Tamura95},
G272.2-3.2; \cite{Harrus01},
G299.2-2.9; \cite{Park07},
RCW 86; \cite{Rho02},
SN 1006; \cite{Yamaguchi08},
Lupus Loop; \cite{Park09},
CTB37A; \cite{Aharonian08},
CTB37B; \cite{Nakamura09},
G349.7+0.2; \cite{Lazendic05},
W28; \cite{KO05},
W44; \cite{KO05},
3C400.2; \cite{Saken95},
Kes 27; \cite{Enoguchi02},
MSH 11-61A; \cite{RP98},
3C391; \cite{KO05},
CTB 1; \cite{RP98},
W51C; \cite{Koo95},
CTA 1; \cite{RP98},
W63; \cite{RP98},
HB21; \cite{RP98},
IC443; \cite{YO09},
Kes 79; \cite{RP98},
HB3; \cite{RP98},
G327.1-1.1; \cite{Sun99},
W49B; \cite{OK09},
3C397; \cite{RP98},
MSH 11-54; \cite{RP98}}
\label{figure6}
\end{figure}

In the present work, we assume that shocks heat substantially ions at the shock front, and then the electron temperature rises through the energy transport by Coulomb collisions and diffusion in the post shock region.  In Figure \ref{figure6} we compare the electron temperature thus obtained \citep{masai94} by the solid lines and temperature in the equipartition with ions at the shock front by the broken lines, with the observed electron temperatures (the circles). One can see the transfer by Coulomb collisions from ions to electrons is a reasonable assumption to account for the observed data. 

\end{document}